\documentclass[aps,prb,twocolumn,amsmath,amssymb]{revtex4}
\usepackage{graphicx}
\usepackage{bm}
\usepackage{color}
\usepackage{multirow}
\usepackage{times}

\begin{document}
\title{Phase diagram of the Cairo pentagonal XXZ spin-1/2 magnet under a magnetic field}
\author{Arnaud Ralko}
\affiliation{
Institut N\'eel, UPR2940, CNRS et Universit\'e de Grenoble, Grenoble, FR-38042
France
} 
\date{\today}
%------------------------------------------------------------
% Abstract
%------------------------------------------------------------
\begin{abstract}
The phase diagram of the XXZ spin-$1/2$ magnet, equivalent to hard-core bosons,
under a staggered magnetic field and on the {\it Cairo} pentagonal lattice is
computed at zero and finite temperature by using a cluster mean field theory
and a stochastic series expansion quantum Monte Carlo.
The complex connectivity and the frustration lead to unconventional 
phases such as a $1/3$-ferrimagnetic plateau stabilized by quantum fluctuations
as well as a $5/12$-{\it topological} phase induced by a local ice-rule
constraint. 
We also report the presence of a ferrimagnetic superfluid and its thermal
melting.
Finally, we discuss the ferro- and antiferro- magnetic (hopping sign) cases.
\end{abstract}
\maketitle

%------------------------------------------------------------
% Text
%------------------------------------------------------------

% Introduction {{{
These last years, important discoveries in strongly correlated physics  have
been reported in systems where frustration plays a central role.
Geometrical frustration is very interesting in that respect. A large variety of
unusual  phases is encountered, from (insulating) exotic spin liquids
\cite{mendels,mila,ran} to superconductivity \cite{johnston}.  Recent
developments on frustrated {\it optical lattices} of cold atoms open 
new directions for stabilizing exotic bosonic phases such as 
supersolids \cite{ruostekoski} or Bose-metals \cite{feigelman,das,paramekanti}.

A case of interest is that of {\it ice-rule} systems for which highly degenerate
classical ground states (GS) are governed by a local constraint \cite{note-MI}.
For spin-1/2 (boson), it corresponds to a fixed number of up (occupied) and
down (empty) spins (sites) on each elementary brick: a tetrahedron on pyrochlore
and checkerboard lattices \cite{harris,ueda} or a triangle on the kagome
\cite{wills}. 
They provide striking features such as charge fractionalization \cite{fulde2},
Coulomb-gas phases and even {\it magnetic monopoles} \cite{book}.  
Their quantum melting is of broad interest but remains a highly non-trivial
question; aside from the strength of the interactions, the lattice geometry can
play a relevant role. Some works in this direction revealed exotic phases,
{\it e.g.} a commensurate {\it Resonating Valence Bond} 
supersolid on the checkerboard lattice~\cite{ralko2}. Hence, it is natural to
search for ice-rule systems with more complex geometries such as inequivalent
site lattices \cite{jagannathan}.

\begin{figure} 
\includegraphics[width=0.45\textwidth,clip]{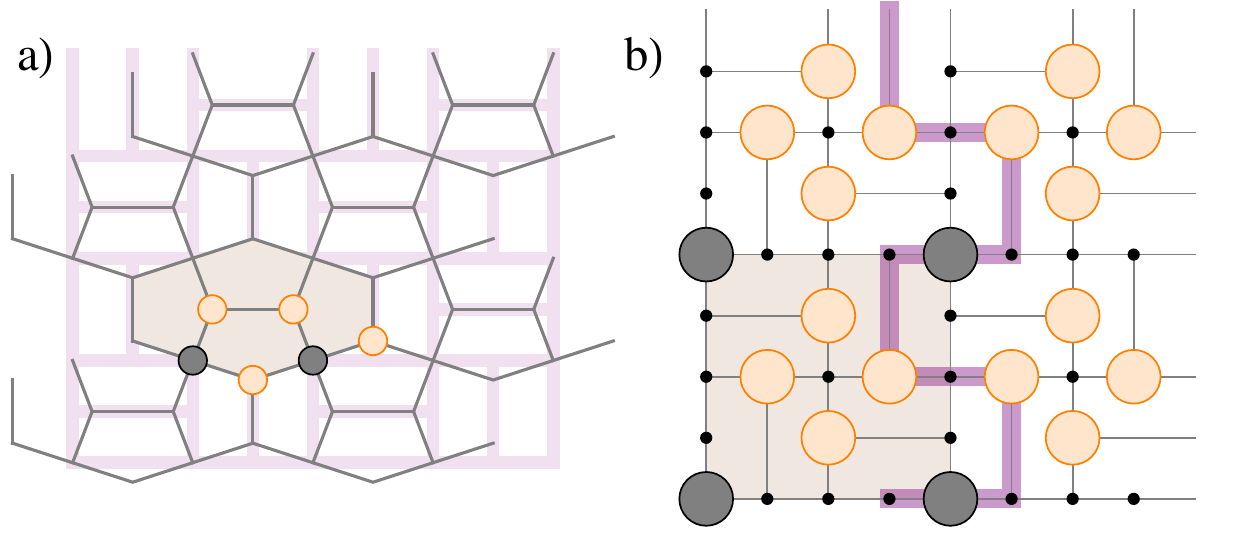}
\caption{(Color online).  a) The {\it Cairo} pentagonal lattice and its 6-site
unit-cell (shaded plaquette) containing $2$ $z_4$ sites (dark points)  and $4$
$z_3$ sites (light points). The continuously deformed square-based  {\it Cairo}
lattice used in this work is displayed in the background.
b) One of the authorized $5/12$ topological {\it ice-rule} configurations with
two spins up (bosons) per pentagon (disks).
Thick line:  typical {\it winding}-loop along which a shift of all encountered
bosons of one lattice site preserves the ice-rule constraint.
Shaded plaquette: the 12-site cluster (2 unit-cells) used in the CMFT.
\label{lattice}
}
\end{figure}

Recently, Ressouche {\it et al.} rendered accessible the two dimensional
{\it Cairo} pentagonal structure  to experiments by proposing the iron-based
compound Bi$_2$Fe$_4$O$_9$ \cite{ressouche}. So far it is the only known
compound with such geometry, a spin-$5/2$ antiferromagnet made of identical
non-regular pentagons with a site dependent connectivity $c_i$ of 3 ($z_3$
site) and 4 ($z_4$ site) neighbors as depicted in Fig.\ref{lattice}-a.  Note
that this {\it Cairo} lattice is the dual of the Shastry-Sutherland lattice for
which the GS could be a spin-liquid \cite{raman}.  
From the  quantum side, substituting the iron atoms by copper ones in the
Bi$_2$Fe$_4$O$_9$\cite{simonet}, or creating complex optical lattices, the
square-based cairo lattice for example (Fig.\ref{lattice}-b), is undoubtably a
major challenge.
However, in view of the recent original phases reported in frustrated systems
\cite{scarola,titvinide,hebert},  inequivalent-site structures possess all the
ingredients to expect unconventional physics.

In this paper, we study a spin-$1/2$ magnet under a staggered magnetic field, or
equivalently the extended hard-core boson Hubbard model, on the {\it Cairo}
pentagonal lattice. 
We report a rich phase diagram obtained both at zero and at
low temperature.
We focus on the different insulating phases: a {\it topological} ice-rule of
two bosons per pentagon, a $2/3$-checkerboard and a pure quantum  $1/3$-{\it
ferrimagnetic} phase with no local constraint.
Moreover,  a large region of ferrimagnetic superfluid is identified as well as
its Kosterlitz-Thouless (KT) transition\cite{kosterlitz} at strong repulsion
induced by thermal fluctuations. 
We also compare the ferro- and antiferro- magnetic cases (sign of the hopping)
and discuss the case of the uniform magnetic field (nature
of the chemical potential).
% }}}

% Model and methods {{{
\section{Model and methods}

Spin-1/2 on the pentagonal lattice can be described
by an extended hard-core boson Hubbard model, with a repulsive nearest neighbor
interaction $V$, a hopping $t$ and a chemical potential $\mu_i$, given by:
\begin{eqnarray}
{\cal H} = - t \sum_{\langle i,j \rangle} (b_i^\dagger b_j + h.c. ) + V
\sum_{\langle i,j \rangle} n_i n_j - \sum_{i} \mu_i n_i 
\label{ham}
\end{eqnarray}
where $i$ is the site index, $b_i^\dagger$ is the creation operator and $n_i$
the number of bosons. 
The correspondence is done by the mapping $S_i^\dagger = b_i^\dagger$ and
$S_i^z = n_i -1/2$ and Eq.(\ref{ham}) is equivalent to a XXZ spin-$1/2$ magnet
with spin couplings $J_z = V$ and $J_\perp = -2 t$, under an effective magnetic
field $h_i$. 
The chemical potential $\mu_i$ is an adjustable parameter \cite{scarola}; this
leads to two important cases for systems with anisotropic $c_i$, it can be
either (a) site dependent or (b) constant. 
In case (a), if $\mu_i$ is set to $\mu +  c_i V /2$, we have the well-known
spin-1/2 magnet under a uniform magnetic field\cite{ioannis}.
In case (b), the system experiences a staggered magnetic field  $h_i= \mu
-c_i V /2$ depending explicitly on the connectivity.
Experimentally, this can occur in systems with  alternating crystal structure
and the possible presence of a Dzyaloshinskii-Moriya interaction\cite{Oshikawa}.
These two cases are of great importance in quantum magnetism, but the latter
case can obviously provide unexpected behaviors as has been recently shown
in 1D materials\cite{Lou}.
In the present work, we hence focus on constant $\mu$ and draw out the rich
phase diagram induced by the complex connectivity $c_i$. We use equivalently
bosonic or spin language when appropriate.

To simplify the analysis, we map the parameters on a sphere described by two
angles $\theta$ and $\phi$ in such a way that  $\bar \mu = \sin \phi$,  $\bar t
= \cos \theta \cos \phi$ and $\bar V = \sin \theta \cos \phi$, where  $(\bar
t,\bar V,\bar \mu)$ are dimensionless. We also consider a deformed {\it
square-based} version of the original lattice, see Fig.\ref{lattice}-b.
\begin{figure}[h] 
\includegraphics[width=0.45\textwidth,clip]{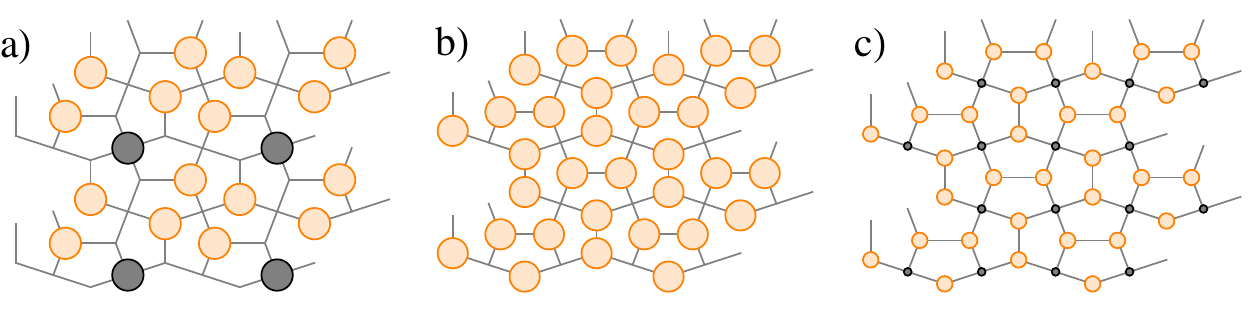}
\caption{(Color online).  Schematic representation of the insulating phases of
Fig.\ref{PhaseDiag}. Dark (light) circles stand for $z_4$ ($z_3$) sites and the
size is proportional to the magnetization (boson density).  (a) A typical
$5/12$-topological ice-rule configuration, (b) the $2/3$-checkerboard and (c)
the quantum $1/3$-ferrimagnetic. \label{Phases}
}
\end{figure}
The phase diagram (Fig.\ref{PhaseDiag}) is obtained by using two numerical
methods: a Cluster Mean Field Theory (CMFT) \cite{zhao}  and a Stochastic
Series Expansion (SSE) Quantum Monte-Carlo (QMC) \cite{sandvik} at
respectively zero and finite temperature.  The basis of the CMFT is to consider
a finite cluster for which internal bonds of Eq.~(\ref{ham}) are treated
exactly whereas the boundary conditions are coupled to an external bath. Here,
we use a 12-site cluster shown in Fig.\ref{lattice}-b (shaded region) and
corresponding to two unit-cells. The system is diagonalized and solved
self-consistently. Note that for the triangular lattice, this method gives an
excellent agreement with the QMC results \cite{hassan}.  Here, it also allows
us to consider the frustrated  $t<0$ case (antiferromagnet) in addition to the
$T$=0 properties. 
The SSE algorithm provides unbiased quantum simulations for very large system
sizes, in our case $N=l \sqrt{3/4} \times l \sqrt{3/4}$ with $l$ up to $72$
(3888 sites), and at finite temperature $T$.
Usually, $T^{-1} = 2l$ is enough to focus on GS properties
\cite{kuroyanagi,Ng}. At very large repulsion however, the Kosterlitz-Thouless
temperature $T_{\textrm{KT}}$ significantly drops down and a finite temperature
transition at small $T$ is expected.
In this work, the phase diagram is computed at  $T^{-1}=100 > 2 l$ (up to
$l=48$) and thermal effects are considered.

We consider four quantities: $x$ the average number of bosons (spin
magnetization), $\rho_S$ the superfluid density (spin stiffness) implemented
via the winding numbers in the SSE \cite{sandvik}, $p_n$ the average number of
pentagons with exactly $n$ up-spins and the order parameter $ M^2({\bf q}) =
\langle \psi_0 | n(-{\bf q}) n({\bf q}) | \psi_0 \rangle / N $ with $n({\bf q})
= \sum_i e^{i {\bf q} \cdot {\bf r}_i} n_i$ performed separately on all sites
of the non-Bravais square-based lattice at ${\bf q} = (\pi,\pi)$
($M_\textrm{all}$) and on the $z_i$ sublattice at ${\bf q} = (0,0)$
($M_{z_i}$)\cite{Ng,remark2}.
The square-based Cairo lattice being a depleted square lattice with extra
bonds ({\it longer} bonds on Fig.\ref{lattice}-b), a finite $M({\bf q})$ is
expected even for a disordered phase, as entirely explained in
[\onlinecite{Ng}].
% }}}

% overview of the phase diagram {{{
\section{Overview of the phase diagram} 
\begin{figure}[h] 
\includegraphics[width=0.45\textwidth,clip]{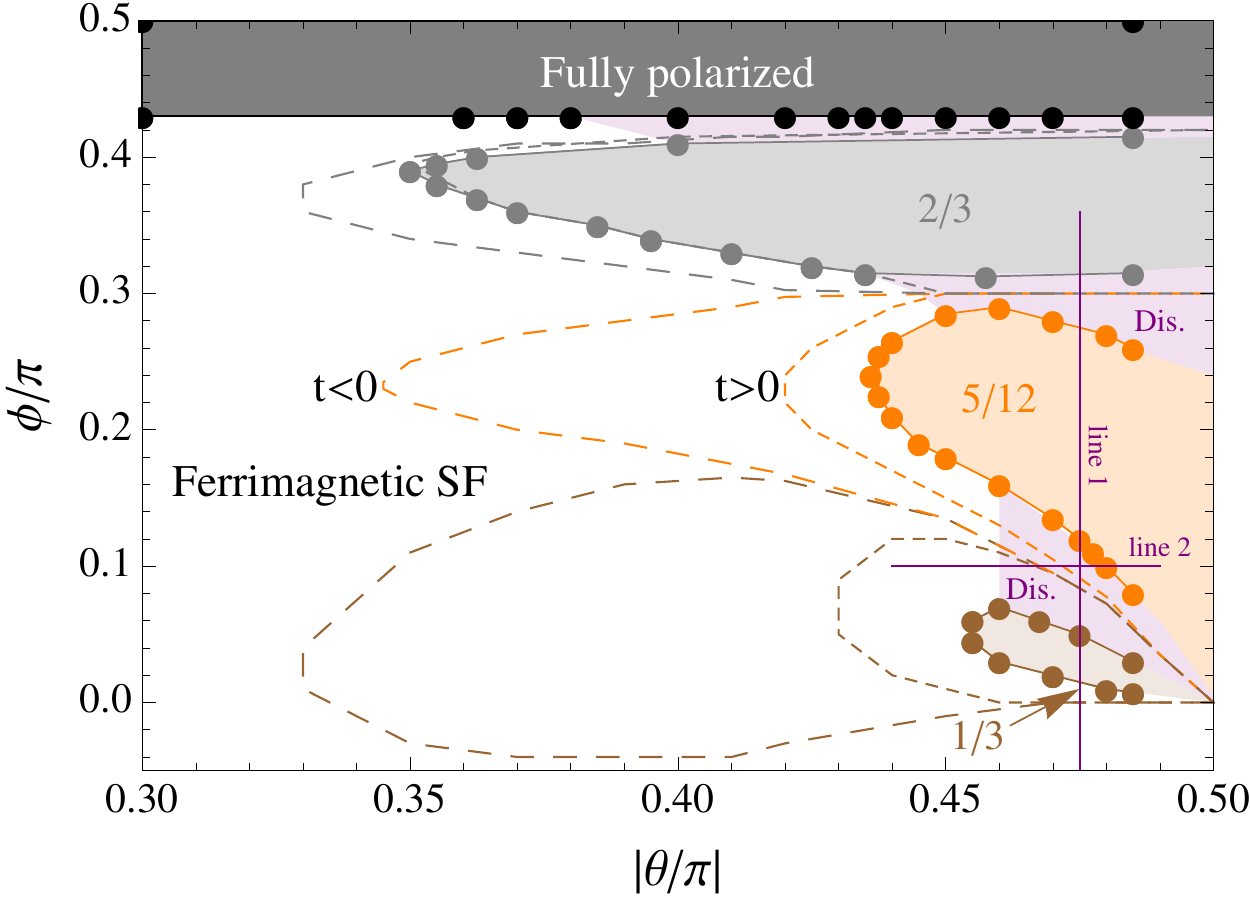} 
\caption{(Color online).  Phase diagram in the $(\theta,\phi)$ plane and the
comparison between zero (CMFT, dashed lines) and finite  (SSE for
$\theta / \pi < 0.485$, symbols) temperature.
Insulating phases (lobes) are enhanced at zero  temperature and by 
frustration ($t<0$).
The ferrimagnetic superfluid (SF) has a Kosterlitz-Thouless (KT) transition to
a disordered region (Dis.) at finite $T$.  Lines 1 and 2 are the scans used
respectively in Fig.\ref{CutTheta} and Fig.\ref{SizeScaling}. 
\label{PhaseDiag}
}
\end{figure}

In Fig.\ref{PhaseDiag} are depicted the
zero (CMFT - dashed lines) and finite (SSE - symbols) temperature phase
diagrams in the large repulsion limit $\theta / \pi > 0.3$ where insulating
phases appear. To characterize the different phases, we have considered two cut
lines respectively at fixed $\theta$ and $\phi$ shown in  Fig.\ref{PhaseDiag}.
Since the phase diagram is very rich, we briefly introduce it in this paragraph
before giving more details in the rest of the paper.
Close to  $\theta=\pi/2$ ($V/t\to \infty$), only reachable by the mean-field,
the frustration leads to  two magnetization plateaus at $x = 2/3$ and $5/12$
(see Fig.\ref{Phases}) at $T=0$. 
Surprisingly, when quantum fluctuations are turned on ($\theta < \pi /2$) a
third insulating plateau at $x=1/3$ arises. This insulator is not stabilized at
the classical limit, is fully driven by the quantum fluctuations and stabilized
by the frustration ($t<0$).
The quantum melting of these lobes leads, in spin language, to a {\it
ferrimagnetic} superfluid (SF) corresponding to different magnetizations
(boson density) on each sublattice.
Uncompensated phases have already been reported in systems with
complex coordination \cite{jagannathan}. 
The finite $T$ phase diagram is computed via the SSE method up to $\theta / \pi
= 0.485$ (circles in Fig.\ref{PhaseDiag}). Stronger interaction results (shaded
regions) are extrapolated.
The main difference with the $T=0$ case is the presence of disordered regions
(Dis.) due to thermal fluctuations, a  Kostertlitz-Thouless transition of
the SF phases and/or the melting of the insulating lobes.  Note that small
discrepancies between the methods cannot be avoided.
In the following, we detail the phases of Fig.\ref{PhaseDiag}. In particular, we
describe the ferrimagnetic character of the SF phase and provide a temperature
analysis at strong interaction before presenting the unconventional 
insulators.
% }}}

% Thermal fluctuations {{{
\section{Superfluidity and thermal fluctuations} 

\begin{figure}[h] \includegraphics[width=0.45\textwidth,clip]{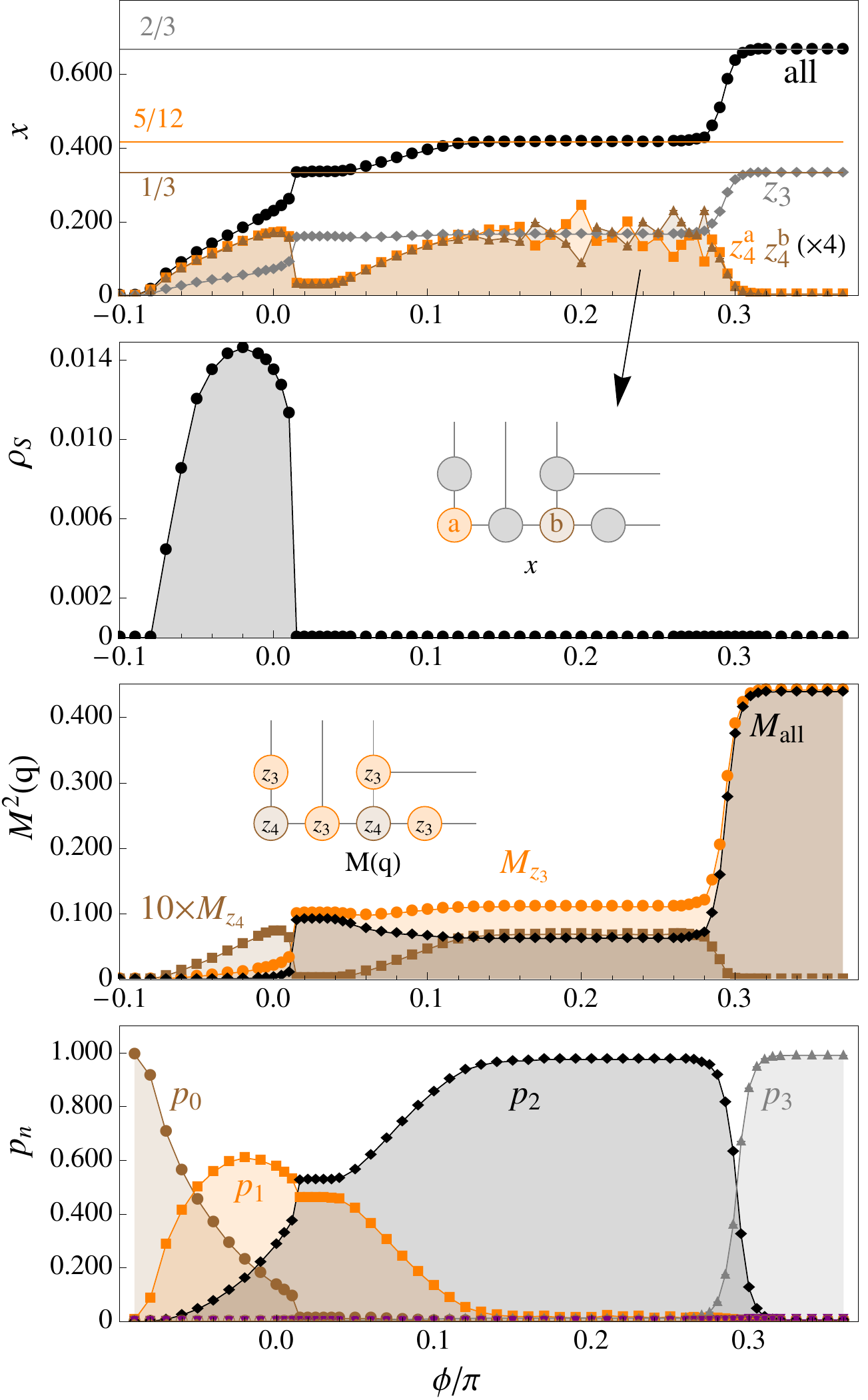}
\caption{(Color online).  Scan along the line 1 of Fig.\ref{PhaseDiag} at
strong repulsion ($\theta = 0.475 \pi$), for the cluster size $l=40$ and at
finite $T$.
From top to bottom: the magnetization $x$ (boson density); the spin
stiffness $\rho_S$ (superfluid density);
the order parameter $M^2(q)$ at $(\pi,\pi)$ for $M_{\textrm{all}}$ (all sites)
and $(0,0)$ for $M_{z_i}$ ($z_i$ sublattice); the average number of pentagons
$p_n$ with $n$ up spins (bosons). 
Inset: sublattices used for $x$ and $M^2(q)$ indicated by arrows.
\label{CutTheta} } 
\end{figure}

\begin{figure}[h] 
\includegraphics[width=0.45\textwidth,clip]{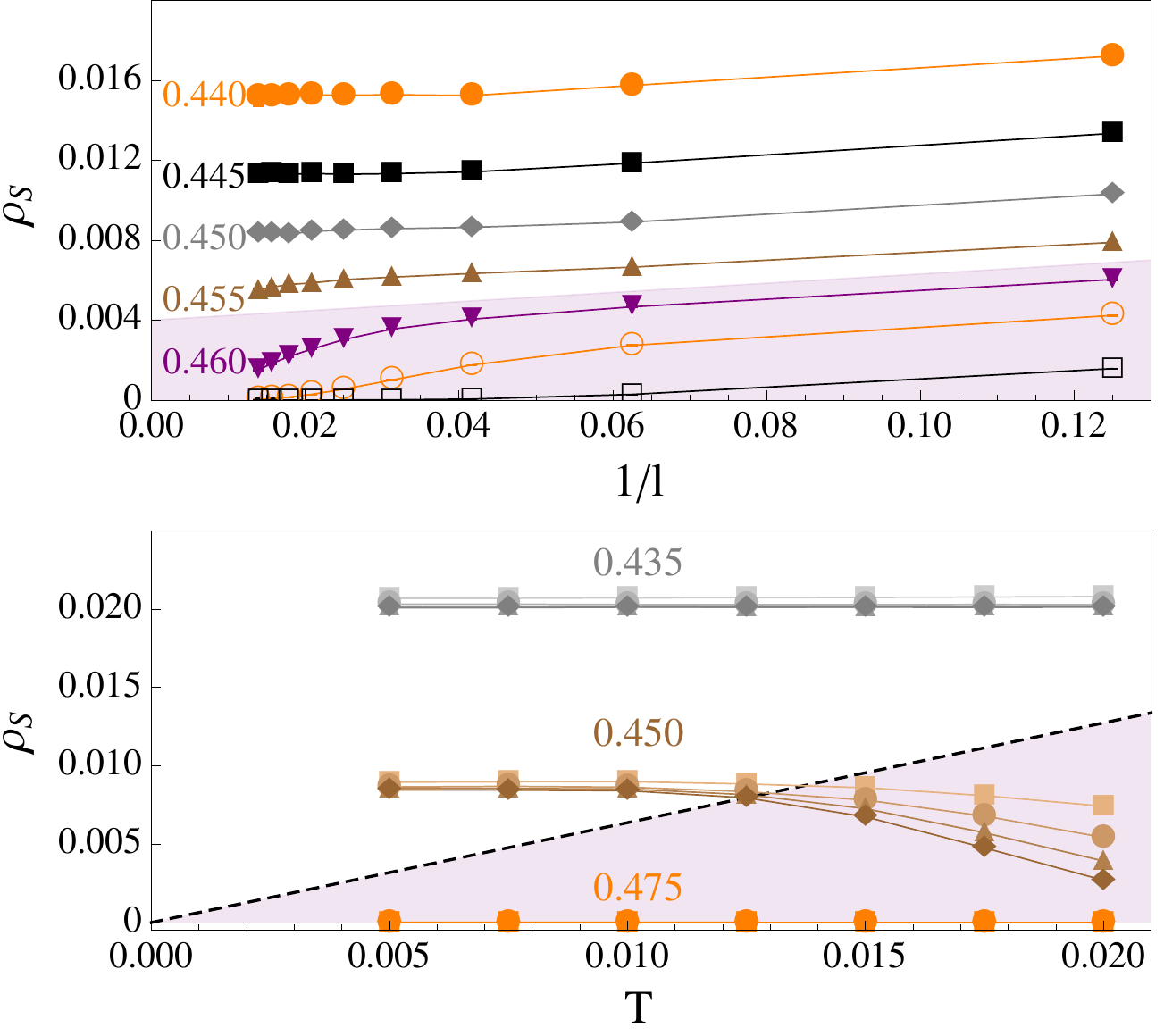}
\caption{(Color online).
Scan along the line 2 of Fig.\ref{PhaseDiag} at $\phi / \pi = 0.1$ of $\rho_S$
as a function of the size $l$ at $T = 0.01$ (upper) and as a function of the
$T$ for sizes up to $l=16, 24, 32$ and $40$ (lower, symbols). 
The dashed line is the $2 T / \pi$ line for which a crossing with
$\rho_S(T)$ indicates the  Kosterlitz-Thouless transition. Shadings:
disordered phases of Fig.\ref{PhaseDiag}.  \label{SizeScaling}
}
\end{figure}

As mentioned above, at zero $T$
and for $\theta $ small enough, the system is a superfluid ($\rho_S \neq 0$)
with an on-site magnetization (boson density) depending on the connectivity
$c_i$. We refer to this phase as the {\it ferrimagnetic} SF\cite{ioannis}. 
We have computed the four physical quantities following two
cut-lines depicted in Fig.\ref{PhaseDiag} at both fixed $\theta$ (line 1,
Fig.\ref{CutTheta}) and $\phi$ (line 2, Fig.\ref{SizeScaling}). 
On Fig.\ref{CutTheta} are displayed from the top to the bottom $x$, $\rho_S$,
$M(q)$ and  $p_n$, as a function of $\phi$ and at finite $T$.
More information can be obtained by computing $x$ and $M(q)$ on the
sublattices $z_3$ and $z_4$ as well. 
For $\phi < 0.2 \pi$, $x_{z_3} \ne x_{z_4}$ while $\rho_S \ne 0$ hence
corroborating the presence of the ferrimagnetic SF. 
At $T=0$ and for $\theta$ large enough  ($>  0.45 \pi$), either transitions of
first order between two lobes or second order with a superfluid are observed
(reinforced at $t<0$).
At finite $T$ (SSE), disordered regions (Dis. in Fig.\ref{PhaseDiag}) with a
finite compressibility $d x / d \phi \ne 0$ and $\rho_S =0$ emerge, related to
a Kosterlitz-Thouless (KT) transition; a finite $\rho_S$ (broken U(1) symmetry)
in 2D is indeed allowed up to a $T_{\textrm{KT}}$ temperature.
In Fig.\ref{SizeScaling} is depicted the size-scaling up to $l=72$ (upper
panel) and the $T$ dependence (lower panel) of  $\rho_S$ for $\phi = 0.1 \pi$
(line 2).
From the upper panel, a transition between $\rho = 0$ and $\ne 0$ occurs at the
thermodynamic limit (TDL).
For 2D systems, the KT transition is located  by the universal jump at
$\rho_S(T_{\textrm{KT}}) = 2 T_{\textrm{KT}} / \pi$ plus some logarithmic
finite size corrections \cite{kuroyanagi}. 
This is calculated in Fig.\ref{SizeScaling} for 3 representative values of
$\theta / \pi$ along line 2: (i) deep in the SF phase ($0.435$), (ii) close to
the transition ($0.45$) and (iii) in the $\rho_S=0$ region ($0.475$).
We obtain (i) $T < T_{\textrm{KT}}$, (ii) $T_{\textrm{KT}} \simeq 0.0125$
close to $T$ and (iii) $T_{\textrm{KT}} \ll T$ with a  KT transition at $\theta
/ \pi \simeq 0.46$.
The finiteness of $T_{\textrm{KT}}$ is non-trivial and beyond the scope of this
paper. However, it is related to the propagation of defects while doping an
insulator. For example, adding a boson in the $5/12$ ice-rule phase results in
creating two $p_3$ defects with zero energy dynamics. This would be compatible
with a small but finite $T_{\textrm{KT}}$ at the TDL. 
% }}}

\section{The incompressible phases}
% The 2/3-phase {{{
{\it The 2/3-checkerboard phase} -- On Fig.\ref{CutTheta}, at $\phi / \pi > 0.3$, all the pentagons
carry  $3$ bosons ($p_3 = 1$), $\rho_S = 0$ and $M(q)$ is
finite for the $z_3$ sublattice while zero on the $z_4$ sites.  This is in
agreement with filled $x_{z_4}$ sites and empty $x_{z_3}$.  This order is a
simple checkerboard crystal (Fig.\ref{Phases}-b) and is the largest lobe of
Fig.\ref{PhaseDiag}, with $\theta_{\textrm{max}} \simeq 0.35 \pi$ at $\phi
\simeq 0.39 \pi$ since adding one particle costs the energy $4 V$. All
the $z_3$ sites are up spins.
% }}}

% The 1/3-phase {{{
{\it The 1/3-{\it ferrimagnetic} phase} --  The $1/3$ plateau is one of the
unexpected phases obtained in this geometry which arises only under quantum
fluctuations (see Fig.\ref{PhaseDiag}), both at zero and finite $T$.  
It is stabilized either under a staggered or a constant\cite{ioannis}
magnetic field and its expansion is 10 times larger (in unit of $V/t$) for
$t<0$. 
It is insulating ($\rho_s = 0$) with no ice-rule constraint ($p_{1,2} \ne 0$,
Fig.\ref{CutTheta}) and no broken lattice symmetry\cite{remark2}. The internal
unit cell densities $x_{z_3}$ and $x_{z_4}$ mismatch (Fig.\ref{CutTheta})
showing a ferrimagnetic character at magnetization $1/3$ (Fig.\ref{Phases}-c).
As displayed in Fig.\ref{CutTheta} at finite $T$, the $1/3$ phase is surrounded
by two phase transitions, a first order with the SF for  $\phi/\pi \simeq 0.01$
and a continuous transition with the disordered phase.
At $T=0$ however, the disordered phase is not present and a first order transition between the $1/3$ phase to the $5/12$ one is obtained (dashed lines in Fig.\ref{PhaseDiag}).
The tip of the $1/3$ lobe in Fig.\ref{PhaseDiag} is located at $\theta / \pi
\simeq 0.455$ precisely where the first order transition vanishes.
% }}}

% The topological ice-rule phase {{{ 
{\it The topological ice-rule phase} -- The $5/12$ plateau is stabilized when the
magnetic field is staggered (constant $\mu_i$). Stable at the Ising limit
(CMFT), this phase is robust against both quantum and thermal fluctuations,
specifically at $t<0$ (Fig.\ref{PhaseDiag}). 
On Fig.\ref{CutTheta},  $\rho_S=0$ and an average of $2$ bosons on $z_3$
sites against $0.5$ on $z_4$ ones per unit-cell is observed. 
$M(q)$ is larger than the depletion induced internal structure signal,
indicating a clear difference between the sublattices\cite{Ng,remark2}.  
With $p_2 = 1$, we deduce the presence of an {\it ice-rule} of two bosons per
pentagon.
A typical configuration of this $5/12$ phase is given in Fig.\ref{Phases}-a.
By labelling the $z_4$ sites inside a unit-cell as $z_4^a$ and $z_4^b$ (inset of
Fig.\ref{CutTheta}), we identify a finite distribution of $x_{z_4^a}$ and
$x_{z_4^b}$ w.r.t. $\phi$ related to the degeneracy of the GS. 
Classical zero-energy configurations in standard ice-rule systems can generally
be connected by quantum tunneling of a finite number of particles on a closed
path. This leads, through perturbation theory, to a quantum effective Hamiltonian
{\it e.g.} quantum dimer models (QDM) \cite{rokhsar,ralko1,ralko2} or loop
models \cite{fulde2,syljuasen}.
Here, no such local moves are available, only {\it winding}-loops invoking the
boundary conditions instead (see Fig.\ref{lattice}-b).
Bosons on such a loop have only two possible positions that respect the
ice-rule constraint and a tunneling from one to the other results in a new
$5/12$ configuration.
For a cluster of size $l$, there are $l/4$ distinct contours in each direction
($x$ and $y$ ).
For a given configuration, the number of bosons on such a contour is a
conserved quantity and each set of these quantities defines a topological
sector. Only the global shift of the bosons along a winding-loop can change
this number and thus the topological sector; the system is protected from local
disorder.
Starting from the most symmetrical $5/12$ configuration (Fig.\ref{lattice}-b),
the total number of winding loops is ${l/2}$. A shift of the bosons along one direction
cancels the possible winding-loops along the other. The number of connected
configurations is then simply $\Omega = 2 \times 2^{l/4}$.
The zero-temperature entropy per site hence scales  as ${\cal S}/N =
\frac{(l+4) \log 2}{3 l^2}$ and vanishes at the $l \to \infty$ limit. 
Since all the configurations are frozen, the phase transition is not smooth, as
confirmed by the sudden appearance of the distribution.
We estimate the width of the $5/12$ plateau by the expansion of this
distribution, {\it e.g.}  $ 0.13(1) \le \phi / \pi \le 0.26(1)$ in
Fig.\ref{CutTheta}.
To our knowledge, isolated sectors have always been reported in systems where
local moves were also available\cite{rokhsar}.
% }}}

% Concluding remark {{{
\section{Concluding remarks} 

We report the phase diagram of spin-$1/2$ magnets
(hard core bosons)  on the {\it Cairo} pentagonal lattice, at zero (CMFT) and
finite (SSE) temperature.
The anisotropic connectivity leads, at constant $\mu_i$, to a staggered
magnetic field\cite{Oshikawa,Lou}.
Various insulating phases are identified among which a pure quantum induced
$1/3$-ferrimagnetic phase, not stabilized in the Ising limit ($\theta = \pi
/2$)\cite{ioannis}.
An original $5/12$-topological ice-rule phase is evidenced, emphasizing the main
difference between spin-$1/2$ systems under staggered and uniform fields.
In this system, the effect of the frustration ($t<0$) enhances the insulating
phases.
Finally, a zero {\it vs.} finite $T$ comparison reveals a KT transition located
at strong repulsion as well as a partial melting of the lobes. 
The two methods employed here are in good agreement.
Open issues remain, such as the complete description of the phase transitions
and the temperature properties.
Nevertheless, the spin-1/2 Cairo magnet is a very promising candidate for
exploring new states of matter.
% }}}

% Acknowledgements {{{
{\it Acknowledgements} -- I thank O. C\'epas and V. Simonet for discussions.
I am indebted to A. La\"uchli and D. Poilblanc for valuable critical
comments of the manuscript. I am also grateful to N. Dempsey for the careful
rereading of this manuscript.
% }}}

% Bibliography {{{

% }}}

\begin{thebibliography}{99}
\bibitem{remark2} No Bragg peak is observed in the structure factor on
the Bravais lattice (6-site unit cell).
\end{thebibliography}
\end{document}